\documentclass{elsart3p}

\usepackage[square,comma]{natbib}

\usepackage{amssymb}
\usepackage{graphicx}

\begin{document}

\begin{frontmatter}



\title{Measurement of sound speed versus depth in Antarctic ice with the South Pole Acoustic Test Setup}
\author{Freija Descamps}
\ead{Freija.Descamps@ugent.be\\}
for the
\author{IceCube collaboration}
\ead[url]{http://icecube.wisc.edu/}
\address{Dept. of Subatomic and Radiation Physics, University of Gent, B-9000 Gent, Belgium}

\begin{abstract}
The feasibility and design of an acoustic neutrino detection array in the South Pole ice depend on the acoustic properties of the ice. The South Pole Acoustic Test Setup (SPATS) was built to evaluate the acoustic characteristics of the ice in the 1 to 100~kHz frequency range. The vertical sound speed profile relates to the level of refraction of the surface noise and determines the reconstruction precision of the neutrino direction. The SPATS speed of sound analysis for pressure and shear waves is presented.
\end{abstract}
\end{frontmatter}

\section{Introduction}
\subsection{Motivation} 
A hybrid optical/radio/acoustic neutrino detector at the South Pole is described and simulated in~\cite{hybrid1,hybrid2}.
The design of an acoustic neutrino detection array as part of this hybrid detector requires knowledge of the acoustic properties of the detector medium. The South Pole Acoustic Test Setup (SPATS)~\cite{Vandenbroucke:2008} was built to evaluate the acoustic attenuation length, background noise level, transient rates and sound speed in the South Pole ice for the 1 to 100~kHz region.
When an ultra-high-energy neutrino interacts in the ice, the hadronic cascade locally and instantaneously warms up a cylindrical portion of the medium followed by a slow dissipation of the heat. The resulting acoustic signature is a pressure wave with a disk-like emission pattern perpendicular to the shower direction. The speed at which this pressure wave then propagates through the South Pole ice is an important factor in event reconstruction and transient background rejection. In addition to obtain knowledge about the absolute value of the sound speed in the ice, it is important to map the vertical sound speed profile. The acoustic waves are bent towards regions of lower propagation speed and the profile dictates the refraction index and the resulting radius of curvature. The emission disk is deformed more for larger sound speed gradients (lower radii of curvature), making the direction reconstruction for the hadronic shower more difficult.

The sound speed depends on the density and temperature of the medium. From the surface to the bulk-ice, in the firn region, the ice becomes gradually stiffer due to the increased density. Therefore the sound speed rapidly increases in this ice layer. A sound speed profile for the firn layer was experimentally obtained by J.G. Weihaupt~\cite{Weihaupt} in 1963 using seismic surface measurements (see also Figure~\ref{fig:3}). 

For larger depths, the density is stable and the sound speed can be modeled using a certain temperature-dependent ice coefficient as input. The temperature of the ice at South Pole has been experimentally determined to go from~-51$^{\circ}$C just below the surface to an expected ~-9$^{\circ}$C at 2800 meters depth, close to the bedrock~\cite{price1}. Therefore, by combining this temperature profile with the sound speed temperature dependence, we can expect a slight decrease of the sound speed towards the deeper ice. For the in-situ measurements presented in~\cite{kohnen}, this temperature dependence of the speed of pressure waves was found to be -2.30 $\pm$~0.17~$\frac{m}{sK}$. In~\cite{Albert} a certain model for the temperature-dependent ice coefficient and J.G. Weihaupt's firn measurements are combined to model the pressure and shear wave speeds for the South Pole ice.
In this model, the surface noise is predicted to be mostly trapped in the refractive firn layer. The human activity on the surface is a likely source of background noise, but the sound should not propagate to the region below firn. A much smaller and downward refraction for the noise originating from the bedrock is expected. 

\subsection{Setup}
The South Pole Acoustic Test Setup consists of 4 acoustic strings. An array of three vertical strings (A, B and C) was deployed in the upper 400~m of the Antarctic ice cap in January 2007. It was extended in the 2007/2008 polar season by the addition of a fourth string (D) down to 500~m depth. Each string was deployed in the upper part of an IceCube~\cite{icecube} hole and has 7 stages with each one transmitter and one sensor module. Strings A, B and C have stages at 80, 100, 140, 190, 250, 320 and 400~m depth while string D has stages at 140, 190, 250, 320, 400, 430 and 500~m depth. This defines 9 distinct instrumented levels. For an overview of the current status of SPATS, see~\cite{Vandenbroucke:2008}. A retrievable transmitter or ``pinger'' was deployed in 6 water-filled IceCube holes. The pinger\footnote{ITC-1001 from the International Transducer Company} offers a broadband omnidirectional emission and has a transmission power of 149 dB per ($\mu$Pa/V) at 1~m distance. It was lowered down to a maximum depth of 500~m while continuously transmitting at 1~Hz. 
\begin{figure}[h]
\centering
\includegraphics[width=7cm,height=6cm]{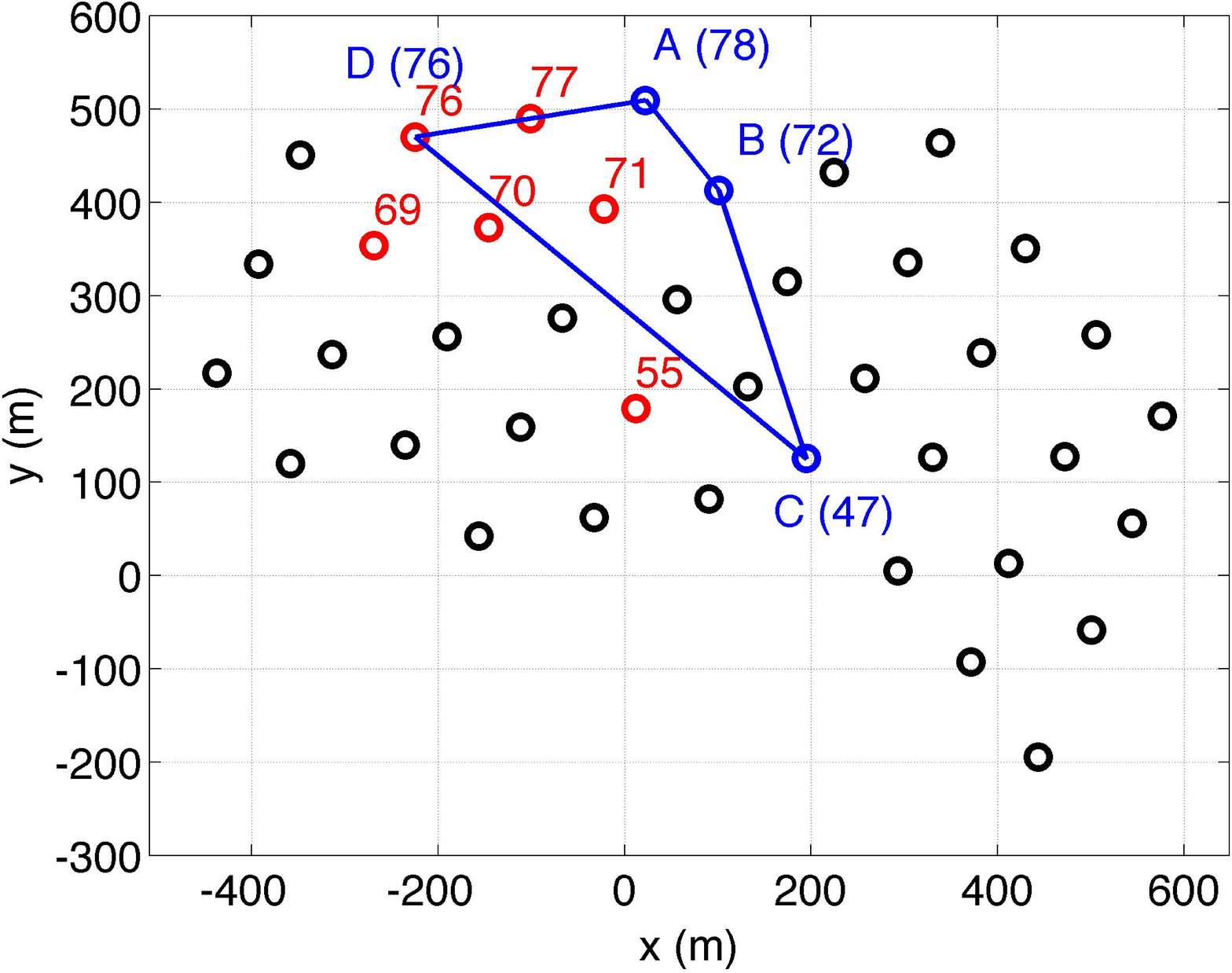}
\caption{Top view of the IceCube detector as of February 2008, the SPATS (47, 72, 76, 78) and pinger (55, 69, 70, 71, 76, 77) holes are labeled. Nominal distance between the nearest-neighbor IceCube holes is 125~m.}
\label{fig:1}
\end{figure}
At each SPATS sensor level, the lowering was stopped for 5 minutes allowing all sensors to record at least 1 waveform of 9 seconds at 200~kHz sampling rate. The analysis presented here uses data from geometries where the pinger and sensor are at the same depth and 125~m apart. Figure 1 shows a schematic of the SPATS and pinger geometry. 
The pinger stage is lowered on a $\sim$2700~m long armored cable through which the power and trigger are routed to the pressure vessel that contains the high voltage (HV) board. Upon receiving the trigger signal, a short ($\sim$60~$\mu$s) HV pulse is sent to the piezo-ceramic transmitter that is suspended about 2~m below the housing. A broadband acoustic pulse is then emitted. Both the SPATS array and the pinger are synchronized by GPS so that the exact transit times of the pinger pulses can be determined. Data were collected with all 9 SPATS instrumented levels. 
An example of a pinger pulse recorded by a SPATS sensor is shown in Figure~\ref{fig:2}.

\section{Sound speed analysis}
\subsection{Method}
The sound speed is expected to be independent of direction in the depths where SPATS modules are deployed. In single-crystal ice the sound speed depends on the direction of propagation relative to the c-axis. But the South Pole ice is polycrystalline, with crystal sizes of order 0.1~cm and with a random distribution of c-axis orientations, such that this effect averages out~\cite{Price}. The frequency dependence of the sound speed is also negligible.
\begin{figure}[h]
 \centering
\includegraphics[width=7.0cm,height=6.5cm]{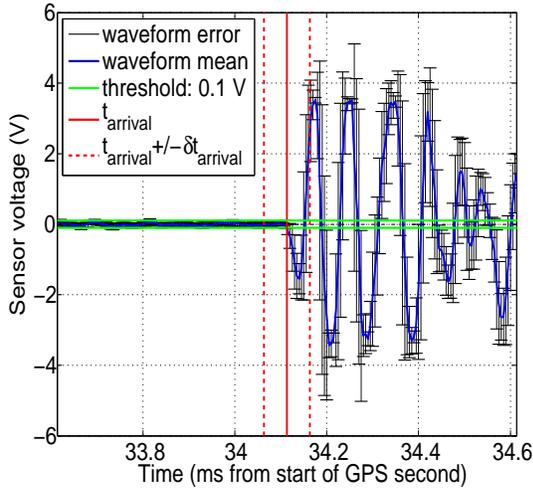}
 \caption{A SPATS sensor waveform and arrival-time (full vertical line) determination using a fixed threshold (full horizontal lines).}
\label{fig:2}
\end{figure}

The SPATS interstring (transmitting with in-ice transmitters and recording on other strings) and pinger data both show shear waves arriving, as expected,  at roughly double the pressure wave transit time for baselines under 200~m. In the case of the SPATS transmitters we can understand the shear waves as being created at the transmitter-ice boundary. As for the pinger emitter, it is located in a column of water where no shear waves can propagate. The observed shear waves are therefore expected to be generated at the water-ice boundary where mode conversion can occur. In this case the relative shear and pressure wave amplitudes depend on the incident angle and therefore on the position of the transmitter in the hole. Arrival times can be extracted from the data for both pressure and shear waves and for all SPATS instrumented levels.

The sound speed for each level is defined as the distance between pinger and SPATS sensor divided by the transit time (time between emission and detection of the pulse). The pinger emits a pulse at every start of a GPS second (1~Hz repetition rate). The SPATS strings are also synchronized with a GPS IRIG-B 100~pps signal so that the start of every second can be determined in the SPATS sensor waveforms. The arrival time ($t_{a}$) is then defined as the time elapsed from the start of the second to the first rising edge of the received pinger waveform. This edge corresponds to the first sample above a threshold that depends on the background noise level of that specific sensor-channel (Figure~\ref{fig:2}). The delay due to cable and electronics ($t_{m}$) is defined at the pinger-side as the delay from the start of the GPS second to the actual acoustic emission. This delay needs to be subtracted from the arrival time in order to obtain the transit time ($t_{t}$) and was measured in a laboratory setup to be 1.9$\pm$0.1~ms. From this we find the transit time: $t_{t}$=$t_{a}$-$t_{m}$.

\subsection{Results}
Three sources of measurement uncertainties are taken into account (see Table 1). First, for the 125~m baseline we find that the error on $t_{m}$ comes to 0.31$\%$ for the 0.1~ms uncertainty. Second, taking a $\pm$0.05~ms window for the arrival time assures a correct selection of the first peak for the waveform; this yields 0.15$\%$. 

\begin{table}[h]
\begin{center}
\begin{tabular}{|c|c|c|c|} 
\hline
Measurement&Result &Error on $t_{t}$\\
\hline 
emission delay & 1.9$\pm$0.1ms&0.31$\%$\\
arrival time &34.00$\pm$0.05 ms&0.15$\%$\\
distance & 125.0$\pm$0.7m& 0.57$\%$ \\
\hline
Total&&0.66$\%$\\
\hline
\end{tabular}
\end{center}
\caption{The measurement uncertainties for a transit time of $\sim$32~ms.} 
\end{table}

Finally, the uncertainty on the horizontal distance between pinger and SPATS sensor needs to take into account the uncertainty on the position of the center of each hole. 
\begin{figure}[b]
\centering
\includegraphics[width=7.5cm,height=6.5cm]{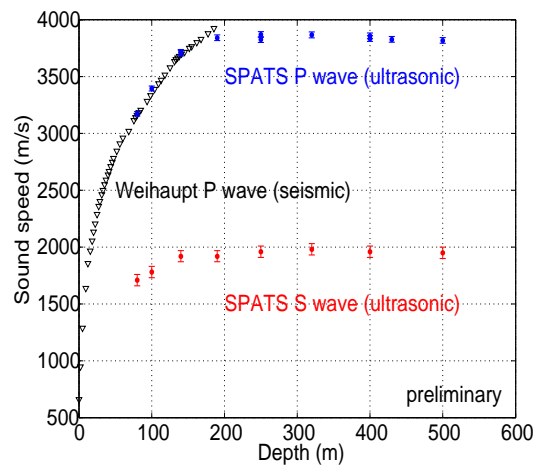}
\caption{v$_{pressure}$ and v$_{shear}$ versus depth from SPATS-pinger speed of sound analysis. 1$\sigma$ error bars are indicated. 
The results from J.G. Weihaupt~\cite{Weihaupt}, for which no uncertainties were cited, are included for comparison.}
\label{fig:3}
\end{figure}
Although the nominal horizontal distance between the centers of two adjacent holes is 125~m, deviations can indeed be expected due to imprecision of initial hole positioning or intentional displacement to avoid surface cabling or debris in the firn. The position of the center of each hole is determined after deployment by a GPS survey with a precision of 0.5~m. From there we assume $\pm$0.5~m in any direction and 0.57$\%$ (0.5$\sqrt{2}$~m for 125~m) as a conservative error on the distance. After combining the three measurement uncertainties, we find the total uncertainty to be 0.66$\%$ for the 125~m baseline. The sound speed analysis can benefit from longer baselines since the error on the distance between transmitter and sensor is dominant. The minimal uncertainty achievable with the 2007-2008 SPATS pinger data is 0.1$\%$ by going to the maximum baseline of 543~m.\newline
Figure 3 shows all SPATS sound speed data points for depths from 80~m to 500~m. There is agreement with~\cite{Weihaupt} in the firn region for the pressure wave measurement.
\begin{figure}[h]
 \centering
 \includegraphics[width=6.5cm,height=5.5cm]{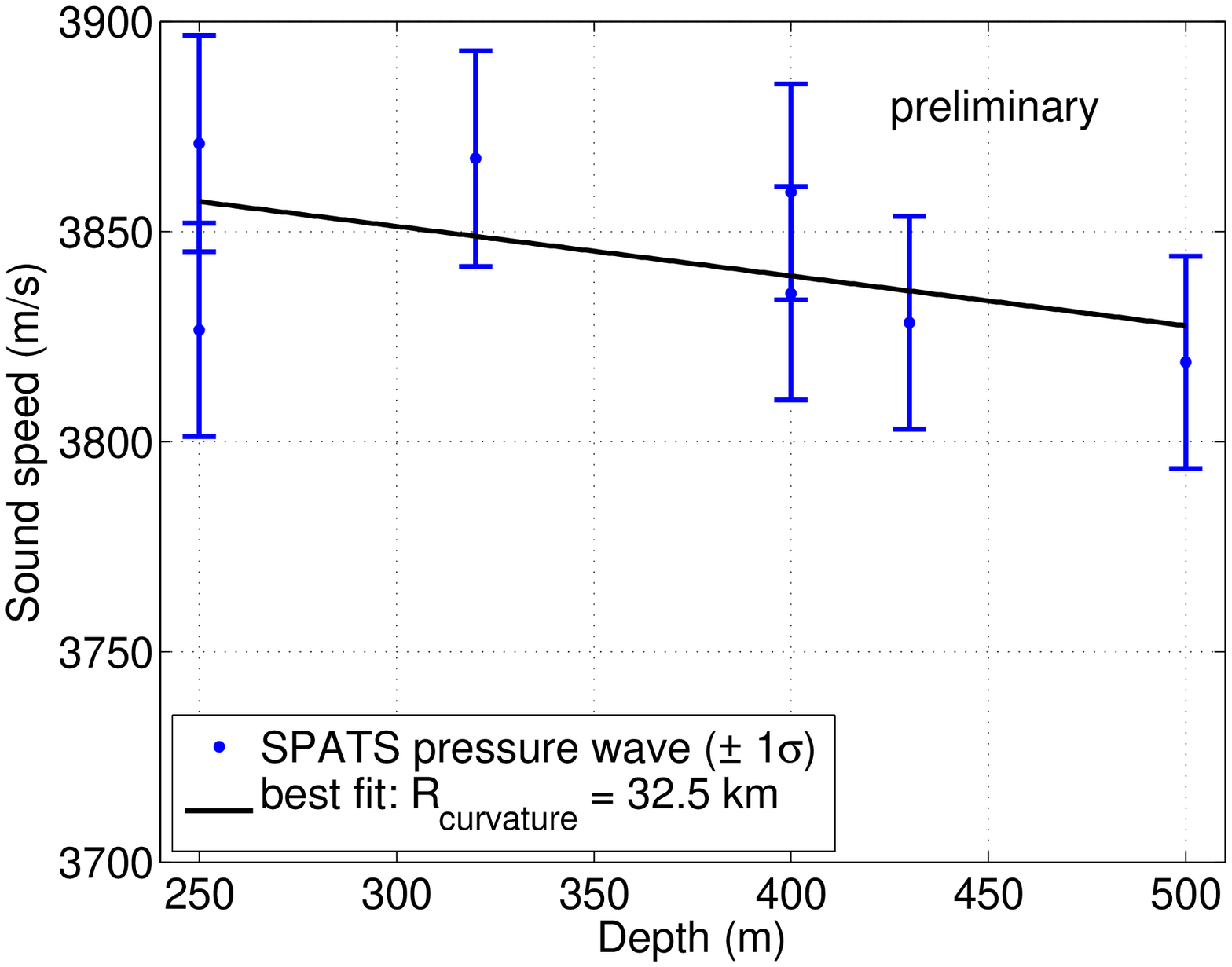}\\
 \includegraphics[width=6.5cm,height=5.5cm]{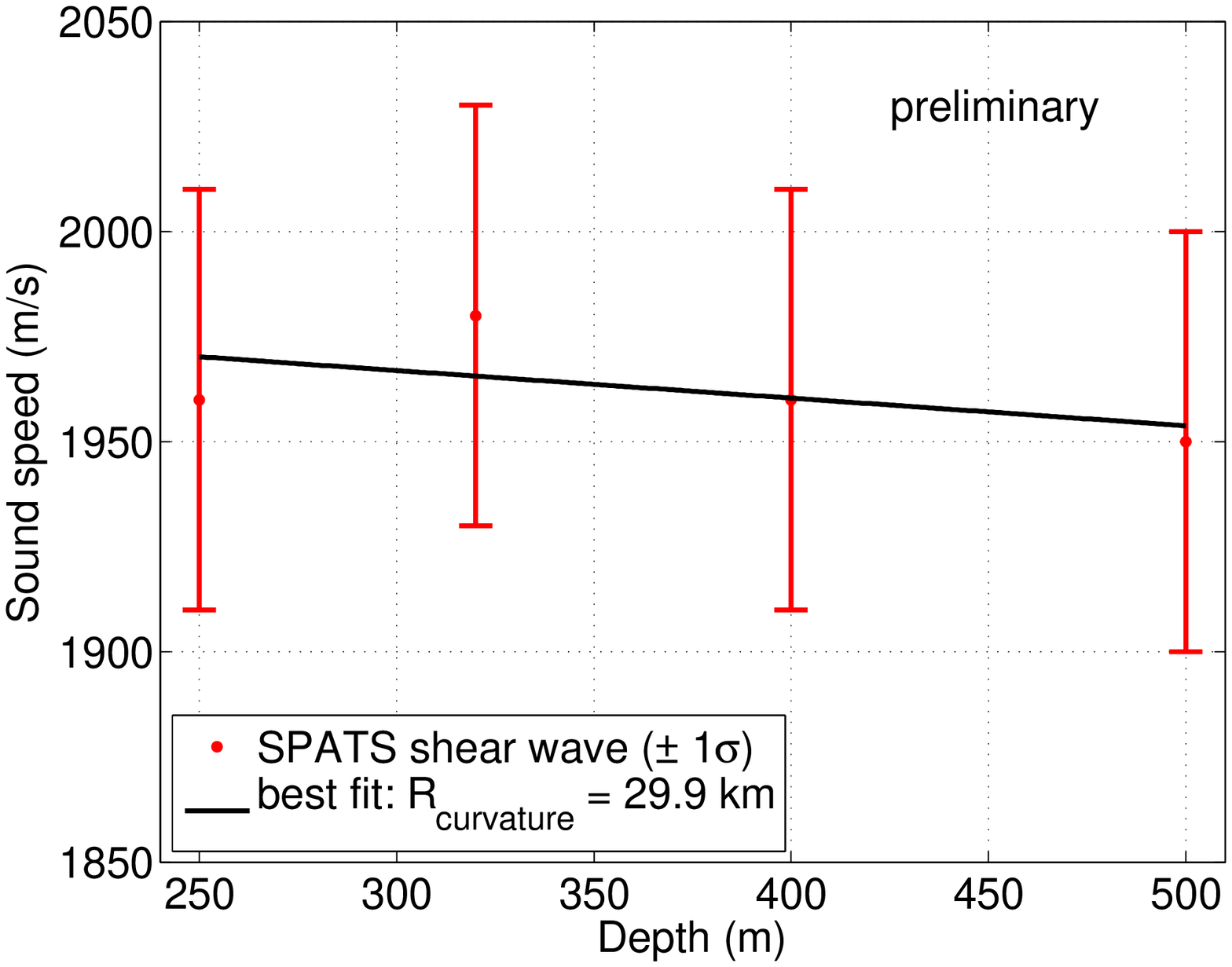}
 \caption{Preliminary results for v$_{pressure}$ (top) and v$_{shear}$ (bottom) versus depth for the bulk ice down to 500~m depth. The best fit is indicated.}
\label{fig:4}
\end{figure}
Figure 4 shows a close-up for the 250 to 500~m depth region for the pressure and shear wave measurements. The graphs include a fit for the sound speed gradient. The best fit yields a radius of curvature of $\sim$ 30~km for both the pressure and shear waves and it is consistent with no refraction. 
Under the hypothesis of no sound speed gradient below 250~m depth, the pressure wave speed was measured to be 3850 $\pm$ 50~m/s and the shear wave speed to be 1950 $\pm$ 50~m/s in the 250 to 500~m depth region. 

\section{Conclusion}
We presented the sound speed analysis and results from the SPATS pinger setup for depths from 80~m to 500~m. Both pressure and shear wave speeds were mapped versus depth in firn and bulk ice. This is the first measurement of the pressure wave speed below 180~m depth in the bulk ice, and the first measurement of the shear wave speed in the South Pole ice. The resulting vertical sound speed gradient for both pressure and shear waves is consistent with no refraction between 250 and 500~m depth.
Currently, an extended analysis is underway that includes longer distances and also a more stringent error estimate for the distance.
\section{Acknowledgment}
We are grateful for the support of the U.S. National Science Foundation and the hospitality of the National Science Foundation Amundsen-Scott South Pole Station..


\end{document}